# Beta frequency shifts in decision making: Spectral fingerprints or communication channels?

Saskia Haegens[1,2*], Julio Rodriguez-Larios[3*], Elie Rassi[4,5*]

[1] Department of Psychiatry, Columbia University, New York, NY, USA

[2] Division of Systems Neuroscience, New York State Psychiatric Institute, New York, NY, USA

[3] Brunel University of London, London, UK

[4] Donders Institute for Brain, Cognition, and Behaviour, Radboud University, Nijmegen, The Netherlands

[5] Department of Psychology and Centre for Cognitive Neuroscience, University of Salzburg, Salzburg, Austria

[*]equal contributions

## Abstract

Recent evidence suggests that beta-band activity plays a key role in decision making. Here we review recent work in humans and non-human primates showing that beta-band frequency shifts in frontal cortex signal categorical decision outcomes. We revisit our previous proposal suggesting that content-specific beta reflects the flexible recruiting of transient neural ensembles and update it to emphasize *frequency* as the relevant parameter. We argue that beta frequency shifts arise from changes in connectivity between weakly coupled oscillators and that, more than a spectral fingerprint, they reflect an active mechanism to (re-)activate behaviorally relevant communication channels in the brain.

## Introduction

While traditionally associated with the sensorimotor system (Pfurtscheller and Silva, 1999; Jones et al., 2009), a growing body of work shows that beta oscillations (~13 – 35 Hz) can be content-specific and reflect the information currently being processed (Spitzer and Haegens, 2017). That is, beta activity allows readout of working-memory content (Spitzer et al., 2010, 2014a, 2014b; Spitzer and Blankenburg, 2011; Salazar et al., 2012; Mendoza-Halliday et al., 2014; Rose et al., 2016; Wimmer et al., 2016) and decision outcomes (Haegens et al., 2011, 2017; Herding et al., 2016; Wimmer et al., 2016; Jayachandran et al., 2023), especially in the context of categorization (Stanley et al., 2018; Wutz et al., 2018a), prior to and independent of translation of such information into a motor response. Critically, these content-specific beta modulations are observed beyond sensorimotor regions, in distributed areas including prefrontal cortex (Spitzer and Haegens, 2017), and likely mediated via thalamocortical interactions (Sherman et al., 2016; Bolkan et al., 2017; Abbas et al., 2018).

We previously proposed that such content-specific beta oscillatory activity reflects the flexible recruiting of transient neural ensembles, e.g., the networks that encode for one decision outcome vs. another (Haegens et al., 2017; Spitzer and Haegens, 2017), but it remained unclear why or how exactly beta activity—typically operationalized as beta-band power or phase coherence—provides this readout of the participant’s decision outcome. Here we revisit that proposed framework in light of recent studies in humans (Rassi et al., 2025) and non-human primates (NHP; Rassi et al., 2023b) showing that the *frequency* of beta oscillations reflects (categorical) decisions. First, we review these findings and discuss their implications and methodological limitations. Then, we update the original framework in light of these new observations and propose a mechanistic explanation of the reported beta frequency shifts based on previous literature and computer simulations. Finally, we broaden the scope to consider other brain rhythms and future directions.

## Evidence: Beta frequency shifts in decision making

### *Beta frequency shifts signal categorical decisions*

Beta peak frequency is highly variable between and within individuals and tasks (Salmelin and Hari, 1994; Baumgarten et al., 2016; Espenhahn et al., 2017). A coarse classification of this variability usually differentiates between low (<20 Hz) and high beta (>20 Hz; Roopun et al., 2006; Kopell et al., 2011; Stanley et al., 2018; Oswal et al., 2021), often conceptualized as distinct sub-bands within the beta frequency range, possibly originating from different (sub-)cortical sources and serving different motor and cognitive functions (Cao et al., 2024; Nougaret et al., 2024). Moment-to-moment modulations of frequency within a particular band have received far less attention (Rassi et al., 2023a) and are the subject of this review.

Kilavik and colleagues (2012) showed systematic modulations of motor beta frequency within beta sub-bands for the first time. During a delay in which monkeys prepared a movement based on a previously presented cue, the direction of the movement being prepared could be decoded from beta frequency. Beyond motor functions, several studies had previously reported that perceptual decisions were reflected in beta power modulations just prior to the decision report (Haegens et al., 2011, 2017; Herding et al., 2016; Wimmer et al., 2016; Stanley et al., 2018). In a recent study in NHP, we show that beta frequency, rather than power, is in fact the key feature: in a series of duration- and distance-categorization tasks in which the boundary between categories changed from one block of trials to the next, beta peak frequency consistently reflected the context-dependent categorical decision, regardless of objective stimulus properties (Rassi et al., 2023b).

In this study, monkeys performed a categorical decision-making task in which they categorized time intervals and distances as either “long” or “short” relative to previously learned categorical boundaries (Mendez et al., 2011; Mendoza et al., 2018; Rassi et al., 2023b; Rodriguez-Larios et al., 2024). Critically, after stimulus presentation, there was a delay in which monkeys could make a categorical decision but not yet indicate it via motor movement, as they did not yet know which motor movement corresponded to their decision. (In the context of decision-making tasks, decisions are often a-priori operationalized as the motor output they produce. One effective way to disentangle decision-related neural activity from movement-related activity is to introduce a decision delay in task designs, after which a response prompt randomly maps decisions to motor responses. By randomizing the response mapping on a trial-by-trial basis and prompting after a decision has been made, one can ensure that a read-out of the decision is independent of subsequent motor activity.) Analysis of local field potential (LFP) recordings in the dorsolateral prefrontal cortex (dlPFC) and pre-supplementary motor area (pre-SMA) during that delay showed that beta frequency predicted the monkey’s decision, independently of the subsequent movement, and independently of the accuracy of the response (**Figure 1A**).

Importantly, the stimuli and categorical boundaries changed from one block to the next (e.g., an interval of 500 ms could be considered short in one block, but long in another), but the same two distinct beta frequencies consistently reflected the two context-dependent categories, regardless of objective stimulus properties. Even when stimuli had identical magnitudes but belonged to different relative categories across task conditions (i.e., depending on the context-defined boundary), beta frequency predicted the animal's response. We conceptualized these two beta frequencies as "channels" of communication, each having distinct spectrotemporal and connectivity profiles. We showed that dlPFC and pre-SMA were connected via these frequency channels, and that these beta dynamics could be characterized as transient bursts (see also Box 1) rather than sustained oscillations. Finally, we showed that the frequency shift was driven by dlPFC, and that category-selective neurons in dlPFC (Mendoza et al., 2018) synchronized with the beta rhythm at the respective category-selective frequency: short-selective cells synchronized with the frequency reflecting the short category, and long-selective cells synchronized with the frequency reflecting the long category (Rassi et al., 2023b).

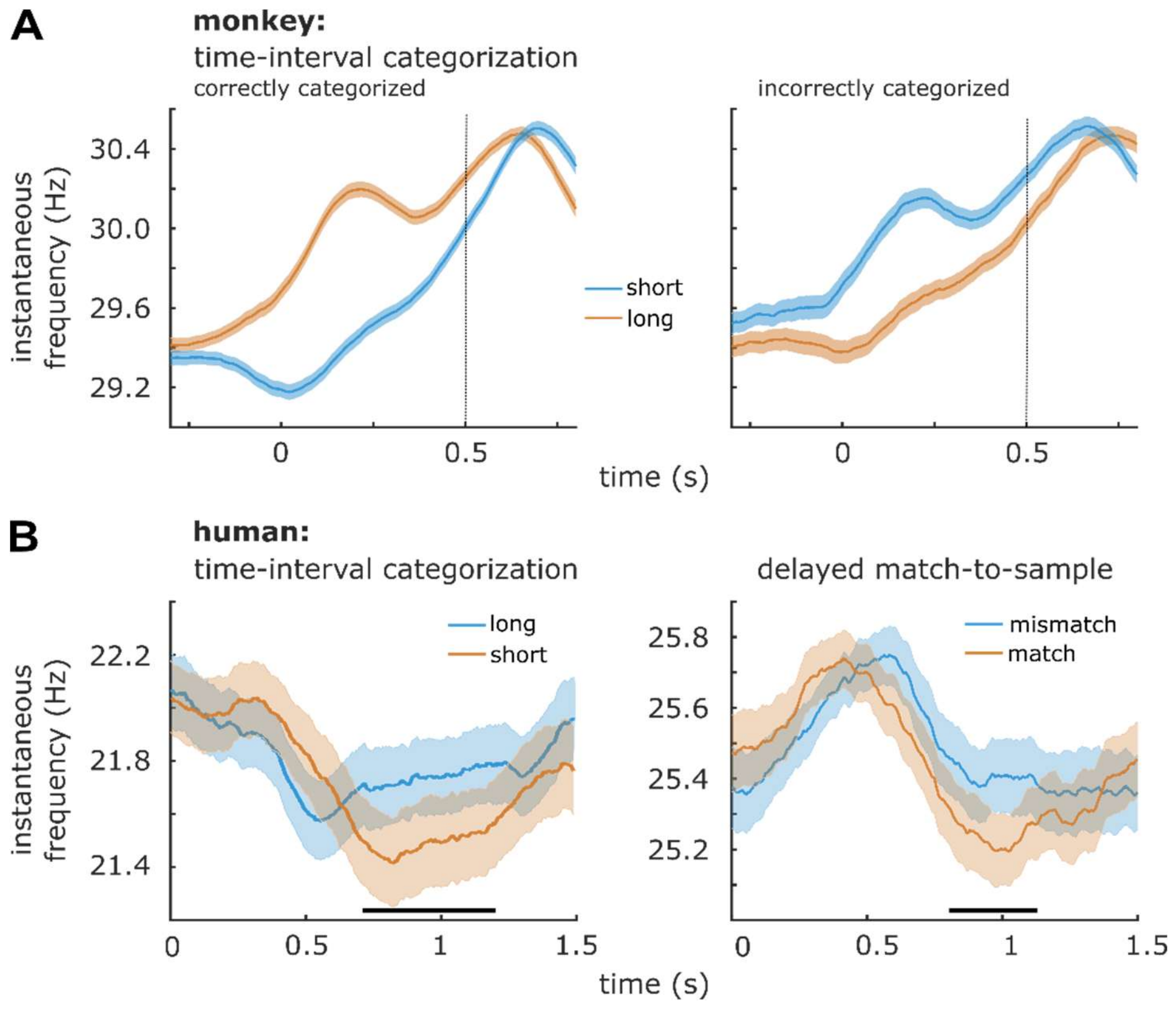

**Figure 1. Beta peak frequency reflects categorical decision outcome in primate dlPFC. (A)** Instantaneous frequency time courses for LFP recorded in monkey dlPFC, for correct (left) and incorrect trials (right). Time zero represents the onset of the decision delay. [Adapted with permission from Rassi et al. (2023b).] **(B)** Left panel: same as A for EEG data localized in human dlPFC, for correct trials. Right panel: same for delayed match-to-sample task. [Adapted with permission from Rassi et al. (2025).]

*Generalization to other paradigms*

Strikingly, we observed a similar pattern of results in human M/EEG recordings across a range of tasks: beta frequency shifts in frontal cortex consistently allowed readout of the subjective decision outcome, independent of physical stimulus properties (Rassi et al., 2025). First, we used an almost identical paradigm as in the NHP study while recording EEG in humans (Rodriguez-Larios et al., 2024). We successfully replicated the main finding that beta frequency in frontal cortex signaled the decision outcome in this categorical decision-making task, although there was more variance in the human EEG data (**Figure 1B**). One caveat is that while our LFP data showed that both monkeys' beta frequency shifted in the same direction to signal long vs. short, our EEG data showed that beta frequency shifted in one direction for two thirds of participants, and in the other direction for the other third. It therefore seems that the direction of the frequency shift is not critical, rather, it is the frequency differential which allows consistent and significant readout of the decision outcome on a trial-by-trial basis within a given participant.

Next, we asked whether this finding would translate to different tasks, sets of stimuli, and recording techniques. We analyzed data from two MEG studies in which participants performed decision tasks (i.e., visual delayed match-to-sample and audio-tactile discrimination) that used response mapping to dissociate decisions from motor outputs. Again, we found that the decision outcome was significantly mapped onto frontal beta frequency—meaning a slower frequency was associated with one decision and a faster frequency associated with another decision—with the direction of the effect again differing across participants. In sum, we find that beta frequency shift in frontal cortex is a decision-related signal that is robust across task designs, decision types, stimuli, analysis approaches, and recording techniques.

## Interpretation: Frequency shifts reflect the recruitment of distinct ensembles

*Frequency as ensemble fingerprint*

Together, our findings suggest that frontal beta sets up flexible neural ensembles at distinct frequencies, which might explain how beta patterns can provide a readout of the information content (**Figure 2**). That is, we propose that beta-frequency dynamics reflect the particular ensemble of neurons that is engaged at a given moment (i.e. the activation of one ensemble vs. another). The specific frequency observed might be a consequence of the particular physiology of the cells in that particular ensemble. This can be thought of in a similar way as the average height of a group of people, randomly selected from a larger population. The average height of the group depends on the individual heights of the people in that sample. In turn, their individual heights depend on many contributing (biophysical) factors. Two randomly selected groups of people will have slightly differing average heights simply because they contain different individuals with different underlying characteristics. In the case of a cell ensemble, the individual cells will additionally influence each other as they interact—in other words, oscillatory frequency is an emergent property of the engaged neural ensemble.

What does this mean? At the bare minimum, it provides us with a spectral "fingerprint" of the particular population of neurons that is engaged (Siegel et al., 2012). In the context of decision making, when two different populations with different spectral fingerprints code for decision A vs. B, this then allows us to read out the decision outcome. In other words, beta-frequency patterns can be used as a proxy for information contained in the underlying population spike activity, providing a powerful biomarker—especially in non-invasive recordings in humans where we do not have access to spikes.

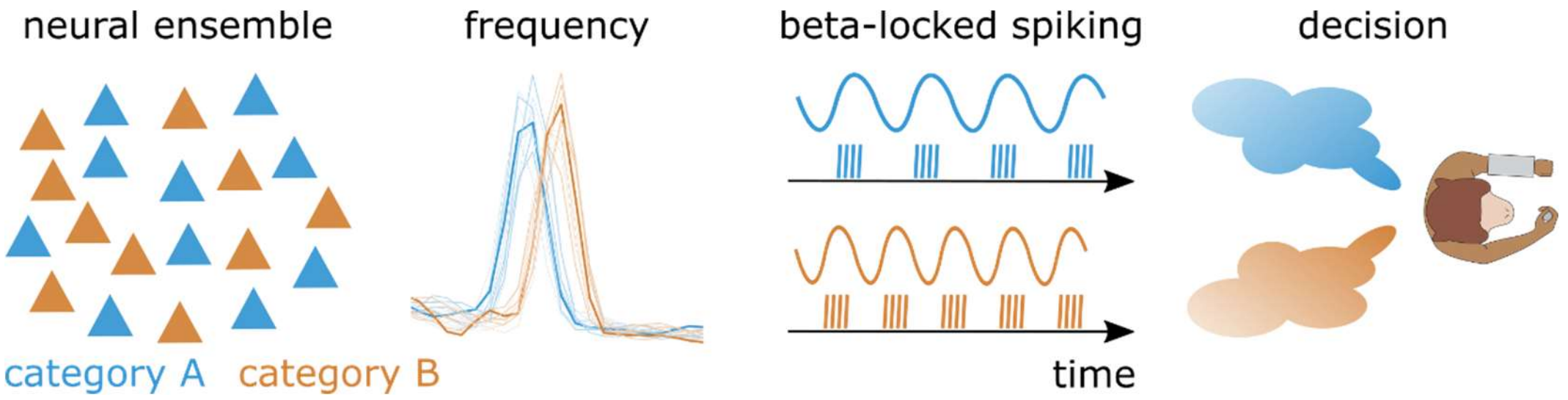

**Figure 2. Neural ensembles selectively signal categorical decisions via distinct beta-frequency channels.** Left: Two overlapping neural ensembles are selective for two different categories A and B. Triangles represent neurons. Center-left: the two neural ensembles have two distinct base beta frequencies. Center-right: Spiking activity of the two neural ensembles is coherent with two different beta frequencies; category information is transmitted downstream via these distinct beta-frequency channels. Vertical lines represent spike trains; waves represent beta rhythms. Right: The two beta rhythms at distinct frequencies provide a readout of the monkey's categorical decision. Clouds represent categorial decisions. [Adapted with permission from Rassi et al. (2023b).]

*Potential biophysical origins of beta frequency variability*

Why do different neural ensembles have different preferred frequencies? It has traditionally been thought that peak frequencies of different rhythms depend on the size of the network involved, with larger networks involving lower frequency rhythms that would be less sensitive to small conduction delays (Kopell et al., 2000; von Stein and Sarnthein, 2000; Buzsáki and Draguhn, 2004). In this view, high frequency rhythms such as gamma are used in local circuits (e.g., within V1), while lower frequency rhythms in the theta/alpha/beta frequency range would be recruited in long/medium-range communication (Kopell et al., 2000). Nonetheless, there are other factors that seem to influence smaller frequency differences within one frequency band, such as receptor expression and synaptic configuration. For example, research on the generation of gamma rhythms has shown that intracellular properties of various cell types (i.e., excitatory and inhibitory cells), their interactions (e.g., recurrent inhibition), network properties, and the involvement of different neurotransmitters/neuromodulators can affect the preferred frequency of a neural population within a narrow frequency range (White et al., 2000; Brunel and Wang, 2003; Buzsáki and Wang, 2012; Fernández-Ruiz et al., 2021; Lowet et al., 2022).

In addition to the size of the network and its properties, recent work has shown that oscillatory frequency also depends on the duration and strength of excitatory input. Specifically, biophysical modelling of beta oscillations suggests that variable frequencies can arise from the integration of synchronous subthreshold excitatory inputs targeting both proximal and distal dendrites of pyramidal neurons (Jones et al., 2009; Sherman et al., 2016; Neymotin et al., 2020; Law et al., 2022). Human Neocortical Neurosolver (HNN) modeling shows that when the distal input is sufficiently strong and sustained for approximately one beta cycle, it can trigger a beta burst. Notably, the duration of this distal drive correlates linearly with burst period, implying an inverse relationship with oscillation frequency. That is, drives with different durations can trigger bursts with different frequencies (Sherman et al., 2016). The ventromedial thalamus—projecting to supragranular layers of the prefrontal cortex (Herkenham, 1980)—has been implicated as a potential source of this distal input, capable of modulating cortical activity without necessarily inducing action potentials (Reichova and Sherman, 2004).

*Frequency shifts and ensemble dynamics*

If neural ensembles vary in oscillatory peak frequency, frequency shifts at the population level then reflect the dynamic changes in ensemble activity. But how are these ensembles formed? Changes in synaptic strength between neurons are likely to play a key role in ensemble formation (Holtmaat and Caroni, 2016). When neurons share a "category receptive field" that favors one category over another, and fire together at the time of categorization, the synaptic strength between them increases. As a consequence, within-ensemble connectivity increases. In this section, we apply the Theory of Weakly Coupled Oscillators (TWCO) and computational simulations to illustrate how changes in connectivity within oscillatory ensembles can shape population-level frequency from a dynamical-systems perspective.

TWCO is a mathematical framework for modelling interactions between oscillators, which has been applied to neural oscillations (Breakspear et al., 2010; Schwemmer and Lewis, 2012; Lowet et al., 2017). This application of TWCO assumes that neural oscillations facilitate communication between different neural populations by the alignment of their excitatory phases, as this would make both spike output and sensitivity to synaptic input coincide in time (Varela et al., 2001). This is in line with prominent theories in the field such as communication through coherence (CTC; Fries, 2015). However, unlike CTC, TWCO conceptualizes synchronization from a dynamical systems perspective—that is, as a non-linear and non-stationary process (Kopell and Ermentrout, 1986; Kuramoto, 1991; Lowet et al., 2022).

In short, TWCO proposes that synchronization involves continuous phase adjustments—through deceleration or acceleration of oscillators—in order to counteract detuning (i.e., frequency difference) and maintain a preferred phase relationship (Lowet et al., 2017, 2022). In this context, two key parameters determine synchronization: i) coupling strength, i.e., the degree of phase adjustments, and ii) detuning, i.e., the mean frequency difference between oscillators. These two parameters can be visualized in the phase response curve, which is obtained by plotting frequency differences as a function of phase differences. The amplitude of this curve determines coupling strength while the mean shows the level of detuning or frequency difference (**Figure 3A**). The effects of the interaction between these two parameters on synchronization are described with an Arnold tongue (**Figure 3B**). In this way, synchronization (as quantified through phase locking value) is maximized in scenarios of low detuning and high coupling strength.

Modulations in the average frequency spectra of weakly coupled oscillators can result from changes in connectivity (Musall et al., 2014). For example, if two subgroups of neurons (with different base frequencies) increase their connectivity in a specific condition, a frequency shift between conditions would be observed in the average population spectrum. This is because the average power spectrum of the area is expected to be dominated by the frequency in which greater overall synchrony occurs. In line with this idea, in vitro experiments in dissociated cortical cultures have shown that as neurons become more coupled, their firing rates converge to a common (average) frequency, which then dominates the network's spectral profile (Penn et al., 2016).

In order to illustrate the proposed principle, we simulated the dynamics of two groups of weakly coupled oscillators (N = 40) in the upper beta frequency range (~25 Hz). Each oscillator has a preferred frequency, and its degree of synchronization with the other oscillators is determined by their detuning and coupling strength (**Figure 3C**). The two groups differ in their average frequency (24 vs. 26 Hz). Signals were generated for 30 sessions (100 trials of 1 second per session) and contained both oscillatory and aperiodic activity (voltage SNR = 2). Connectivity was modulated in a subgroup of oscillators by either increasing the coupling strength or decreasing the frequency detuning. Common detuning and coupling strength between all oscillators results in an average population frequency spectrum with a peak at 25 Hz (**Figure 3D**), i.e., the average of the two groups. In contrast, when connectivity is increased in a subgroup of oscillators by either increasing coupling strength (**Figure 3EG**) or decreasing frequency detuning (**Figure 3FH**), a frequency shift can be observed in the average spectrum. Specifically, the main frequency of the spectrum of all oscillators depends on the average frequency of the subgroup with higher connectivity (**Figure 3E-H**). In other words, if a subgroup of oscillators with a relatively higher peak frequency increases its connectivity, the average frequency spectrum will show an increase in its peak frequency.

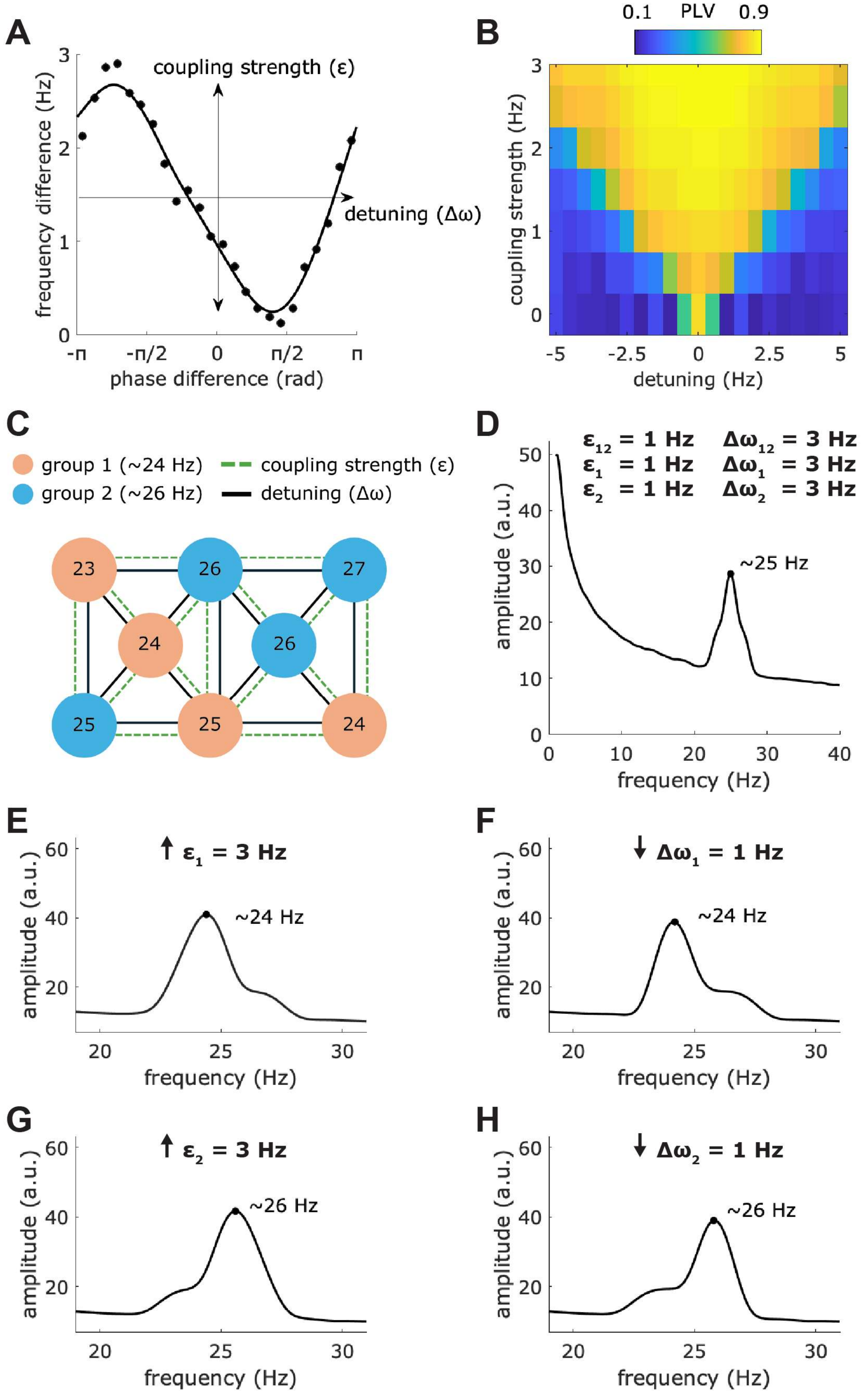
A
coupling strength (ε)
detuning (Δω)
frequency difference (Hz)
phase difference (rad)
B
0.1 PLV 0.9
coupling strength (Hz)
detuning (Hz)
C
group 1 (~24 Hz)
group 2 (~26 Hz)
coupling strength (ε)
detuning (Δω)
23 26 27 24 26 25 25 24
D
ε12 = 1 Hz Δω12 = 3 Hz
ε1 = 1 Hz Δω1 = 3 Hz
ε2 = 1 Hz Δω2 = 3 Hz
~25 Hz
amplitude (a.u.)
frequency (Hz)
E
ε1 = 3 Hz
~24 Hz
F
Δω1 = 1 Hz
~24 Hz
G
ε2 = 3 Hz
~26 Hz
H
Δω2 = 1 Hz
~26 Hz

**Figure 3. The effect of connectivity modulation in the spectrum of weakly coupled oscillators. (A)** Phase response curve of two weakly coupled oscillators containing coupling strength and detuning parameters. **(B)** Arnold tongue describing the relationship between phase locking (PLV), interaction strength, and detuning. **(C)** Depiction of simulation. Two groups of weakly coupled oscillators with predefined base frequency and pairwise connectivity (derived from coupling strength and detuning parameters). **(D)** Average population power spectrum of simulated oscillators during baseline (with equal detuning and coupling strength across groups). **(E-H)** Changes in population peak frequency when connectivity is increased in a subgroup of oscillators (by either increasing coupling strength or decreasing detuning in that subgroup), showing that the peak frequency of the average spectrum corresponds to the average frequency of the subgroup of oscillators with highest connectivity.

### *Frequency as communication channel*

Beyond a fingerprint reflecting which neuronal population is synchronously active, we propose that beta oscillations provide distinct frequency channels, allowing for the selective transmission of information to downstream regions. This is in line with what Akam and Kullman refer to as frequency-division multiplexing (Akam and Kullmann, 2010, 2014). In this model, oscillatory frequency can be conceptualized as metadata to distinguish signals within a multiplexed system.

Multiplexing is the process of integrating multiple signals for transmission, enabling the distinct parts to be individually recovered afterward. In this view, information content is encoded at the level of population (spike) activity, while oscillations at a particular frequency serve as a channel to selectively transmit the code downstream, where a network with the appropriate filter settings can selectively read out the transmitted information (Akam and Kullmann, 2010). Hence, the downstream read-out depends on the filtering properties of the downstream region. While receiving multiple inputs at multiple frequencies, a subpopulation of the downstream region can selectively read out the input whose frequency it is selective to. In this way, information can be routed through frequency-selective filtering rather than requiring separate (anatomical) pathways. In this framework, “filtering” refers to frequency-selective responsiveness (e.g., through resonant network dynamics or subthreshold resonance), rather than requiring the receiving population itself to oscillate at that frequency.

More generally, selective communication does not strictly require matching intrinsic properties between sending and receiving ensembles. Rather, communication depends on the interaction between the dynamically emerging frequency of the sending population and the frequency-selective filtering properties of the receiving population. Which sending frequencies are present depends on the neuronal ensembles recruited by the current task, whereas which frequencies can be read out depends on the filtering properties of downstream subpopulations. For example, if after completing a task where the categories are A and B, a new task is introduced where the categories are C and D, subpopulations of neurons that code for C and D might be different than those coding for A and B, and their respective preferred frequencies may or may not overlap in the sender region. Consequently, the frequency channels used for communication can change dynamically as different neuronal ensembles are recruited, without requiring a fixed one-to-one correspondence between specific categories and specific frequencies across tasks.

Transient oscillatory bursts at distinct frequencies, such as observed in the beta band (Jones, 2016; Sherman et al., 2016; Rodriguez-Larios and Haegens, 2023; Lundqvist et al., 2024; Rayson et al., 2026), are particularly well-suited for such a mechanism, as they can open a communication channel not only at a specific frequency but also at a specific time (in the case of our studies, during the decision delay) (Akam and Kullmann, 2014; Kim et al., 2026). Networks participating in a burst of synchrony at a particular frequency may therefore temporarily increase their effective connectivity with each other relative to networks bursting at different frequencies or different times. This idea is in line with the suggestion by Lundqvist et al. (2024) that “beta bursts may act as spatiotemporal filters that modulate information flow on a brain-wide scale”.

The idea of neural oscillations providing distinct frequency channels for neural ensembles recently received support from recordings in rats during spatial/object learning (Fernández-Ruiz et al., 2021). In short, this study showed that different neurons in entorhinal cortex synchronized with different parts of the dentate gyrus at distinct gamma frequencies during spatial navigation vs. object recognition tasks. Based on these findings, the authors argue that oscillatory frequency could segregate neuronal messages, and this way facilitate a target downstream (“reader”) area to disentangle convergent inputs.

## Outlook

### *Frequency shift as emergent property*

We here considered models on three different levels to interpret our frontal beta frequency shift observations: i) at the neural circuit level, showing how frequency variability can emerge as a consequence of different biophysical factors involved in beta generation, such as receptor expression, synaptic configuration, and/or excitatory drive (Buzsáki and Wang, 2012; Sherman et al., 2016), ii) at the ensemble level, showing how frequency shifts might emerge from the interactions between neurons that behave like weakly coupled oscillators (Lowet et al., 2017), and iii) at the network level, showing how frequency shifts might facilitate information transmission between areas by using oscillatory frequency as ‘’metadata’’ to distinguish signals within a multiplexed system (Akam and Kullmann, 2010, 2014).

In our view, beta oscillations are an emergent property of the engaged neural ensemble, and moment-to-moment changes in frequency directly reflect the engagement of a different ensemble. We suggest that the specific frequency of a particular ensemble—which can be measured on the population level—does not only depend on the biophysical properties of its constituent cells and their driving inputs, but also on their coupling dynamics. Together, we propose that, rather than a simple spectral fingerprint, beta oscillations provide distinct frequency channels that allow selective transmission of information to downstream regions.

### *Do frequency shifts reflect changes in excitability?*

In the above interpretation, we assume that frequency shifts reflect the activation of different neural ensembles. Alternatively, it has been suggested that frequency shifts do not so much reflect the activation of a different population, but rather, a change in the level of excitability of a given population, thereby modulating spike thresholds and spike timing variability (Cohen, 2014; Mierau et al., 2017). Cohen (2014) suggests that “frequency sliding can be a means of modulating neural excitability—a gain-control mechanism.” Strictly speaking, on the level of our current LFP observations (let alone M/EEG), we cannot determine whether a shift in frequency reflects a change within the same group of neurons, or a shift in activation from one population to another. However, a shift between populations seems most plausible, as i) it is not as obvious how to explain the frequency difference between two categorical decision outcomes in terms of excitability levels within one population, and ii) our single-unit spike results, where we can identify whether a particular cell codes for decision A vs. B and whether it preferentially locks to frequency A vs. B, point towards this being two sets of cells operating at two slightly different frequencies (Rassi et al., 2023b). It is also possible that these two mechanisms are complementary and that they occur in different contexts. In this regard, future studies using invasive recordings with high-density arrays (Tchoe et al., 2022; Palopoli-Trojani et al., 2024) are needed to conclusively assess whether the reported frequency shifts occur within the same neural population or reflect the activation of different subpopulations. Such experiments could expand on our NHP observation and concurrently record a large number of category-selective cells to establish whether cells encoding the same category/decision indeed form ensembles synchronized at specific beta frequencies.

*Are frequency shifts cause or consequence of neural communication?*

In line with prominent theories in the field (Varela et al., 2001; Fries, 2005, 2015), we assume that phase synchrony between neural populations and communication are tightly linked. However, our model predicts that frequency shifts emerge because of the differential phase synchronization of different subpopulations of neurons with slightly different resonant frequencies, with the most synchronized subpopulation dominating the power spectrum. It follows that coherence between the brain region that contains category-sensitive neurons (e.g., dlPFC) and a downstream region (e.g., preSMA) will peak at the dominant frequency, as shown in our NHP study (Rassi et al., 2023b). Importantly, frequency modulation enables dynamic network reconfiguration by shaping transient synchronization states between neural populations. In this view, frequency is an emergent property of the engaged neural ensemble (and not “top-down” imposed), and dynamic rather than stable and fixed. The latter is in line with the empirical observation that beta dynamics are burst-like (Jones, 2016). Finally, our proposal is agnostic as to whether frequency (or phase synchrony) is a cause or consequence of communication (Schneider et al., 2021). In order to test the latter, a causal link between beta oscillatory frequency and behavior needs to be assessed.

*Causal role*

Neuromodulation techniques could be used to test a putative causal link between beta oscillations and decision making. For example, “artificial” beta oscillatory electrical fields could be created in the brain non-invasively through transcranial alternating current stimulation (tACS; Riddle and Frohlich, 2021) in order to assess whether this biases decision-making performance. Although one previous study showed that beta tACS to frontal cortex selectively affects decision making in the context of a temporal categorization task (Wiener et al., 2018), a causal link between specific decisions and beta frequencies has not yet been established. Based on our previous findings, we predict that stimulating at the (subject-specific) beta frequency associated with decision A vs. B would cause behavioral biases towards the respective decision. If done in animal models, this would additionally allow assessing how stimulation affects spiking of cells encoding decision A vs. cells encoding decision B (Krause et al., 2019).

Since prior work has shown frontal beta to also play an important role in decision making in rodents (Bolkan et al., 2017; Symanski et al., 2022), optogenetic neuromodulation combined with in-vivo invasive electrophysiological recordings could provide further causal evidence (Fernández-Ruiz et al., 2021; Ibarra-Lecue et al., 2022). Optogenetic techniques could be used to introduce, inhibit, or modulate specific beta frequencies in targeted neuronal populations and evaluate if this biases decision outcomes accordingly. Crucially, this approach could also give us insights about the biophysical origins of frontal beta oscillations (e.g., which types of neurons are necessary to generate the rhythm). For example, one could manipulate the duration of the distal dendritic input in prefrontal neurons, which is thought to determine beta oscillatory frequency (Sherman et al., 2016), and assess the consequences for neural oscillations and subsequent behavior. Combined with high-density recordings, this would further allow assessing whether frequency shifts reflect excitability changes within a given population or the switching between different populations.

Another avenue for future work is in patient populations with idiosyncratic beta patterns and potentially related cognitive deficits, e.g., in Parkinson's disease (Little and Brown, 2014) and schizophrenia (Uhlhaas and Singer, 2010). Neuropharmacological manipulations could be used to study impact on beta frequency dynamics and decision-making behavior (Iskhakova et al., 2021). The here proposed neuromodulation approaches could further benefit from biophysical modeling to develop mechanistic hypotheses, such as made possible by the HNN model, which can then be validated and constrained with empirical work (Tolley et al., 2026).

*Open questions*

The evidence discussed here focused primarily on (categorical) decision-making tasks with binary decisions. An obvious question is whether our observations hold for cases with more than two possible decision outcomes, that is, does the framework generalize beyond binary categorical decisions? We propose to address this question by extending the experiment to include a second task with three categories: short, middle, long, in addition to the original two-category task. One potential outcome is that (for each individual participant) the new "middle" category maps onto a frequency in between the original frequencies associated with "short" and "long" decisions, while the short and long decision outcomes remain reflected by the same peak frequencies (i.e., the same recruited neural ensembles) as in the two-category task. For example, if in the two-category task the "short" ensemble has a peak frequency at 20 Hz, and the "long" ensemble at 26 Hz, the "middle" ensemble might end up at 23 Hz. A possible interpretation of such an outcome is that the middle category is represented by an ensemble that has some overlap with both the original (short and long) ensembles, resulting in a "middle" frequency. An obvious follow-up question is then whether the original bandwidth defines the limits of the available frequency space, or whether additional categories could recruit frequencies beyond those observed in the original two-category task. Another possible outcome of the three-category experiment is that the added "middle" category ends up represented by a new ensemble, with a frequency that is unrelated to (and not necessarily in the middle of) the original two. By extension, it is possible that in this scenario the original two categories (short and long) also become (re)mapped onto new ensembles with different frequencies. A possible interpretation of this potential outcome is that each version of the categorization task recruits a different subset of neurons from a larger population, such that the mapping between categories and ensemble frequencies is re-established dynamically for each task**.** Finally, it is also possible that increasing the number of categories eventually reduces the separation between ensemble frequencies to the point where decision outcomes can no longer be distinguished based on peak frequency. This could reflect either limitations in the signal-to-noise ratio of the recorded signals or a limitation of the proposed framework itself.

A related question is how a particular frequency becomes "associated" with a particular category/decision outcome. We suggest tackling these questions in future experiments by observing how frequency shifts come about during learning and in response to rule updating. We predict that during learning, specific decision outcomes become "mapped" onto particular frequencies by virtue of the cell ensemble that dynamically forms to represent that particular decision (Antzoulatos and Miller, 2014; Stanley et al., 2018). The particular frequency can then be understood as an emergent property of that neural ensemble, and the neural ensemble akin to a "decision receptive field". The emergence of separable peak frequencies corresponding to specific decision outcomes should then coincide with task acquisition. Combining this with the multi-category condition, one could introduce a cue indicating on a trial-by-trial basis which rule needs to be applied (e.g., two- vs. three-way categorization) and ask if the same cells represent the respective decision outcomes and to what extent ensemble participation dynamically switches and on what timescales. Such dynamics could be conceptually akin to representational drift, where neuronal ensembles dynamically change over time due to spontaneous, self-organized synaptic plasticity while maintaining stable memory representations (Kalle Kossio et al., 2021).

To be clear, we do not propose that there are (hard-coded or otherwise) separate frequencies for all conceivable categories or decisions. The specific neural ensemble generating a specific beta rhythm during any given observation is likely the result of a dynamic process, and in a way arbitrary (that is, there is no meaning to one category being reflected by the higher vs. lower beta rhythm; see also the interindividual differences in the observed frequency patterns). Rather, we propose that these dynamic subpopulations can flexibly map (and remap) onto particular categories relevant for the current task, temporarily synchronizing at a particular beta frequency that we can use to read out the categorical decision outcome—the core question is what neurophysiological mechanism underlies the observed beta frequency dynamics. Another open question is whether the proposed mechanism generalizes beyond the types of 2AFC decision tasks discussed here, and more broadly, whether this mechanism applies to maintenance of information beyond decision outcomes.

*Beyond frontal beta frequency shifts*

Based on our prior results, we here focused on frontal beta dynamics in the context of (categorical) decision making. However, the proposed framework could be applied to other brain rhythms originating from different networks and relevant to other cognitive functions. In fact, frequency shifts have previously been reported in the theta (Axmacher et al., 2010; Rodriguez-Larios and Alaerts, 2019; Senoussi et al., 2022), alpha (Haegens et al., 2014; Samaha and Postle, 2015; Wutz et al., 2018b), and gamma bands (Lowet et al., 2015).

Prior literature shows that frequency shifts in different brain rhythms are behaviorally relevant in different tasks. In rodents, the frequency of the hippocampal theta rhythm has been consistently linked with running speed (Bland and Vanderwolf, 1972; McFarland et al., 1975). More recent work in humans has shown that higher cognitive demands are associated with increases in alpha frequency and decreases in theta frequency (Haegens et al., 2014; Rodriguez-Larios and Alaerts, 2019; Senoussi et al., 2022; Rodriguez-Larios et al., 2026). In the same line, there is substantial evidence showing that the peak frequency of alpha oscillations in visual cortex affects the temporal resolution of visual perception, favoring visual segregation at higher frequencies and visual integration at lower frequencies (Samaha and Postle, 2015; Wutz et al., 2018b).

Could frequency shifts in different rhythms and networks be generated by the same mechanisms? Recent literature shows that neural oscillations behave like weakly coupled oscillators in both local circuits (e.g., gamma in V1; Lowet et al., 2017) and distributed networks (e.g., fronto-parietal theta-alpha rhythms; Rodriguez-Larios et al., 2026). Hence, it is possible that frequency shifts in different frequency bands and networks are reflective of changes in connectivity of neural populations with different preferred frequencies.

*Conclusion*

We here reviewed recent evidence from studies in humans and NHP showing that frontal beta frequency shifts signal categorical decisions. We propose that the observed frequency modulations emerge from the recruitment of distinct neural ensembles when different (categorical) decisions are made. Specifically, we argue that frontal beta frequency shifts result from changes in connectivity between subgroups of weakly coupled oscillators with slightly different resonant frequencies. Beta frequency shifts can then be understood as the activation of behaviorally relevant communication channels, allowing for the selective transmission of information. The proposed mechanism might apply to brain rhythms beyond the beta band, as well as to other cognitive contexts. Future studies combining high-density electrophysiological recordings and neuromodulation techniques (e.g., in rodents) are needed to further uncover the neurophysiological origins and computational principles of behaviorally relevant frequency modulations.

**BOX 1: Detecting genuine beta oscillations**

The detection and quantification of genuine beta oscillations in M/EEG recordings is not trivial. During wakefulness, beta oscillations (13 – 35 Hz) are significantly less prominent (i.e., showing less amplitude and duration) than oscillations in lower frequencies such as alpha (8 – 13 Hz) and theta (4 – 8 Hz) (Klimesch, 1999; Sherman et al., 2016). Critically, beta rhythms are difficult to disentangle in the frequency domain because brain rhythms in the theta-alpha range have non-sinusoidal properties, that affect the beta frequency range (Schaworonkow, 2023). This is because Fourier-based methods typically used in neuroscience assume a sinusoidal basis and decompose non-sinusoidal waveforms into sums of sinusoids at roughly harmonically related frequencies (Cole and Voytek, 2017). Consequently, non-sinusoidal rhythms in the theta-alpha frequency range tend to show spectral peaks in the beta range (**Figure 4**). Two examples of such non-sinusoidal rhythms are: i) the somatosensory mu rhythm, with a main peak around 10 Hz and a second harmonic around 20 Hz (Kulhman, 1978; Schaworonkow, 2023), and ii) the frontal sawtooth theta rhythm, with a main peak around 5 Hz and a third harmonic around 15 Hz (Onton et al., 2005).

The confounding effect of non-sinusoidal rhythms on analysis of M/EEG beta dynamics can be mitigated in several ways. One approach is to use spatial filters to isolate brain areas dominated by beta oscillations (Nikulin et al., 2011; Grandchamp et al., 2012; Bonaiuto et al., 2021). In support of this approach, invasive electrophysiology has shown that in areas such as motor cortex, oscillatory activity is most prominent in the beta range (Stolk et al., 2019; West et al., 2023). However, it is difficult to completely rule out the influence of other rhythms when investigating areas with a broader frequency profile (i.e., areas that oscillate at different frequencies in a task-dependent manner; Keitel and Gross, 2016). Another possibility is to restrict the analysis to periods of high beta activity, thereby making it less likely to be artifactually caused by other rhythms. As activity in the beta range has been shown to appear transiently or burst-like (Jones, 2016), different "beta burst" detection algorithms have been developed (Sherman et al., 2016; Bonaiuto et al., 2021; Enz et al., 2021; Szul et al., 2023). Chiefly, these algorithms select time periods in which beta power surpasses a specific threshold. Importantly, this approach also allows to distinguish oscillatory properties that are normally invisible to non-time-resolved spectral measures, such as peak-to-peak amplitude, rate, coverage and waveform shape. In this regard, we have recently developed a beta burst detection algorithm that further minimizes the possible influence of non-sinusoidal theta-alpha rhythms by excluding time periods in which beta activity coincides in time and space with lower frequency rhythms with a relatively higher amplitude (Rodriguez-Larios and Haegens, 2023).

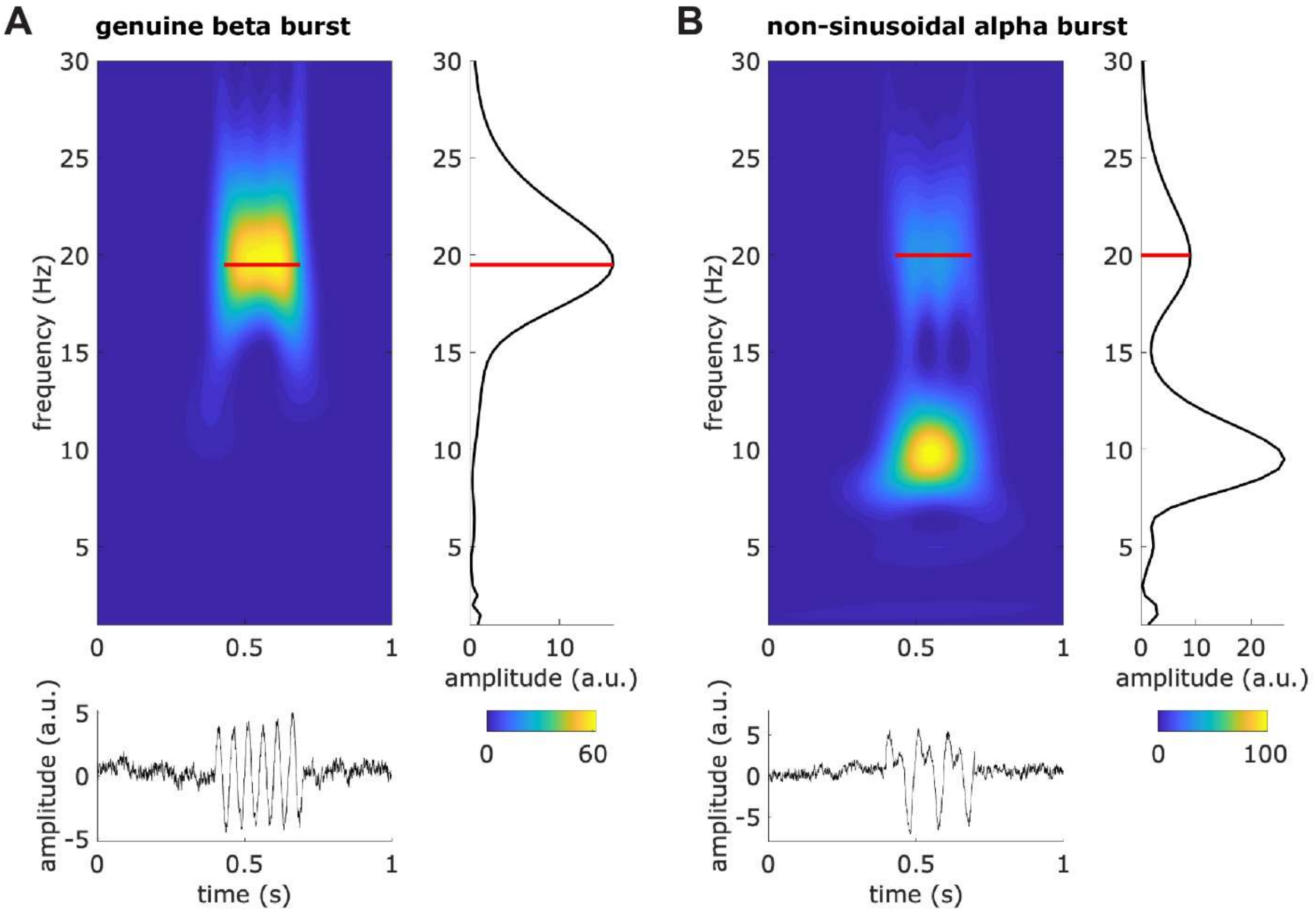

**Figure 4. Genuine versus spurious beta bursts. A)** Time-frequency representation (top left) and power spectrum (top right) of a beta burst around 20 Hz along with its time-domain representation (bottom). Simulated data; red line indicates peak beta frequency. **B)** Same for an artifactual beta burst. In this case, the power increase in the beta range (~20 Hz) is due to the non-sinusoidal properties of an alpha burst occurring around 10 Hz.

**BOX 2: Methods to detect frequency shifts**

The main frequency of an oscillation can be extracted in different ways. The simplest and most commonly used method involves extracting the local maximum of the Fourier-derived power spectrum in a pre-defined frequency band. When a single local maximum cannot be identified, main frequency can also be estimated using the *center-of-gravity approach*, which consists of a power-weighted average of the frequency band of interest (Klimesch, 1999). More recently, Cohen (2014)–introduced the *frequency sliding method*, which estimates the instantaneous frequency by taking the temporal derivative of the unwrapped phase angle obtained from the analytic signal (via the Hilbert transform) after band-pass filtering. Although this latter method provides better temporal resolution than spectrum-based methods, the frequency estimates are less precise as they entail median filtering to minimize artefacts caused by phase slips (Cohen, 2014).

Both spectrum- and Hilbert-based methods used to estimate the main frequency of neural oscillations are susceptible to changes in aperiodic activity. Aperiodic activity refers to the non-oscillatory component of the signal, which is commonly referred to as 1/f as it decreases exponentially in power as a function of frequency (Donoghue et al., 2022). Several methods have recently been developed to control for the aperiodic component of the signal. Several of these methods directly parametrize the aperiodic contribution and subtract it from the signal in the frequency domain (Whitten et al., 2011; Donoghue et al., 2020; Samaha and Cohen, 2022; Seymour et al., 2022). Alternatively, the aperiodic contribution can also be estimated non-parametrically by performing an irregular resampling of the signal (Wen and Liu, 2016).

Another important factor that affects the frequency estimation of a neural oscillator is the definition of frequency bands. Methods relying on bandpass filtering and Hilbert transform (Cohen, 2014) are more affected by the specific frequency band definition than spectrum-based methods that focus on the maximum peak frequency at specific time points (Wilson et al., 2022). In this regard, several methods have been developed to estimate frequency bands in a data-driven manner (Watrous and Buchanan, 2020; Cohen, 2021). An alternative approach is to use Empirical Mode Decomposition (EMD) instead of bandpass filtering prior to frequency estimation (Huang et al., 1998; Fabus et al., 2021; Quinn et al., 2021). However, it is important to note that although EMD can effectively isolate the oscillation of interest without a priori definition of frequency bands (and without assuming sinusoidal basis), it can also split it into different components (i.e., "mode mixing"; Fabus et al., 2021). Another strategy for frequency estimation that does not require a priori frequency band definition consists of directly detecting peaks and troughs in the time domain (Cole and Voytek, 2019), though this method might only be adequate for neural oscillations with a relatively high signal-to-noise ratio.

Finally, it is important to underline that observable changes in peak frequency could be due to: i) an oscillator actually changing its frequency, or ii) a different oscillator originating in a different area becoming more strongly observable in the signal (Rodriguez-Larios et al., 2022; Bilgen et al., 2026). To assess this possibility, spatially distinct neural oscillations need to be disentangled before peak frequency estimation. For that purpose, blind source separation techniques such as independent component analysis (Delorme et al., 2012) and spatio-spectral decomposition (Nikulin et al., 2011) can be used (potentially in combination with dipole fitting to estimate their brain sources (Delorme and Makeig, 2004)), as well as source separation techniques that take into account a priori knowledge of brain oscillatory activity (Meij et al., 2016; Fulvio et al., 2024). Spatial source reconstruction techniques, such as *minimum norm estimation* (MNE; Hämäläinen and Ilmoniemi, 1994) and *beamforming* (Van Veen et al., 1997; Westner et al., 2022), can be used to disentangle different brain rhythms by reconstructing signals into virtual source channels after incorporating a priori knowledge of underlying anatomy and conductivity distribution. More recently, *Hidden Markov Models* have also emerged as a viable alternative to identify transient and spatially localized brain rhythms in a purely data-driven manner (Vidaurre et al., 2016, 2018; Zhou et al., 2025).

## Acknowledgements

E.R. is supported by the Austrian Science Fund (FWF) Erwin Schrödinger Fellowship J4580 and NWO OCENW.XL21.XL21.069 (SoMeMe). S.H. is supported by NIH R01 MH123679 and NSF CRCNS 2424100. The authors would like to thank Andreas Wutz for helpful comments on the manuscript.

**Declaration of interests**

The authors declare no competing interests.

**Author contributions**

The three authors contributed equally to the work.